\documentclass[aps,twocolumn,showpacs,pra]{revtex4}

\usepackage{exscale}
\usepackage{graphicx}
\usepackage{amsmath}
\usepackage{latexsym}
\usepackage{amsfonts}
\usepackage{amssymb}
\usepackage{amscd}
\usepackage{bbm}
\usepackage{dcolumn}      
\usepackage{bm}           

\usepackage{pstricks,pst-grad,fancybox,graphics}
\usepackage{enumerate}

\newcommand{\bra}[1]{\ensuremath{\langle#1|}}

\newcommand{\ket}[1]{\ensuremath{|#1\rangle}}

\newcommand{\ketbra}[1]{\ensuremath{| #1 \rangle \langle #1 |}}

\newcommand{\BE}{\begin{equation}}
\newcommand{\EE}{\end{equation}}
\newcommand{\be}{\begin{equation}}
\newcommand{\ee}{\end{equation}}
\newcommand{\ben}{\begin{displaymath}}
\newcommand{\een}{\end{displaymath}}
\newcommand{\bea}{\begin{eqnarray}}
\newcommand{\eea}{\end{eqnarray}}
\newcommand{\bean}{\begin{eqnarray*}}
\newcommand{\eean}{\end{eqnarray*}}
\newcommand{\kommentar}[1]{}

\newcommand{\mean}[1]{\ensuremath{\langle #1 \rangle}}

\newcommand{\proj}[1]{\ketbra{#1}}
\newcommand{\tr}{{\rm Tr}}
\newcommand{\diag}{{\rm diag}}

\newcommand{\bc}{\begin{center}}
\newcommand{\ec}{\end{center}}
\newcommand{\proofend}{\hfill\fbox\\\medskip }

\newcommand{\Eins}{\ensuremath{\openone}}
\newcommand{\eins}{\ensuremath{\openone}}

\begin{document}

\title{Not all pure entangled states are useful for sub shot-noise interferometry}

\author{Philipp Hyllus$^1$, Otfried G\"uhne$^{2,3}$, 
and  Augusto Smerzi$^1$}
\affiliation{$^1$BEC-CNR-INFM and Dipartimento di Fisica, Universit{\`a} di Trento,
I-38050 Povo, Italy\\
$^2$Institut f{\"u}r Quantenoptik und Quanteninformation, {\"O}sterreichische Akademie der Wissenschaften,
Technikerstra{\ss}e 21A,
A-6020 Innsbruck, Austria\\
$^3$Institut f{\"u}r Theoretische Physik, Universit{\"a}t Innsbruck, 
Technikerstra{\ss}e 25, A-6020 Innsbruck, Austria}

\date{\today}

\begin{abstract}
We investigate the connection between the shot-noise limit in
linear interferometers and particle entanglement. In particular, we ask 
whether or not sub shot-noise sensitivity can be reached with 
all pure entangled input states of $N$ particles 
if they can be optimized with local operations.
Results on the optimal local transformations allow us to show that 
for $N=2$ all pure entangled 
states can be made useful for sub shot-noise interferometry 
while for $N>2$ this is not the case. 
We completely classify the useful entangled 
states available in a bosonic two-mode interferometer.
We apply our results to several states,
in particular to multi-particle
singlet states 
and to cluster states. The latter turn out to be practically 
useless for sub shot-noise interferometry.
Our results are based on the Cramer-Rao bound and the 
Fisher information.

\end{abstract}


\maketitle

\section{Introduction}

The field of quantum interferometry has received much attention recently
due to the prospect of enabling phase sensitivities below the shot-noise, 
with applications in various fields such as
quantum frequency standards, quantum lithography, quantum positioning
and clock synchronization, and quantum imaging \cite{GiovannettiSci04}.
Current research on linear interferometers 
is directed at the search for optimal input states and output measurements
\cite{CavesPRD81,YurkePRA86,HollandPRL93,DowlingPRA98,SoerensenNat01,
CamposPRA03,BollingerPRA(R)96,PezzePRA(R)06,MorePezze,UysPRA07,CablePRL10},
adaptive phase measurement schemes \cite{BerryPRL00,HigginsNat07,BerryPRA09,HuangPRL08}, 
and the influence of particle losses \cite{DornerPRL09,DemkowiczPRA09,RosenkranzPRA09}.
Several proof-of-principle experiments reaching a sub shot-noise
sensitivity have been performed, for a fixed number of particles with photons 
\cite{MitchellNat04,WaltherNat04,NagataSci07,MatthewsNatPhot09,GaoNPhys10} and
ions \cite{LeibfriedNat05}, while squeezed states for interferometry 
with a non-fixed number of particles have been prepared 
with Bose-Einstein-condensates \cite{OrzelScience01,JoPRL07,EsteveNat08,GrossNat10,RiedelNat10},
atoms at room temperature \cite{FernholzPRL08} and light \cite{GodaNatPhys08,VahlbruchPRL08}. 
Also schemes for non-linear interferometers are under investigation
\cite{LuisPLA04,RoyPRL08,BoixoPRL07,BoixoPRL08,ChoiPRA08}.

In this article, we are interested in the connection between 
particle entanglement and phase estimation for linear interferometers 
with input states of a fixed number of particles $N$. 
It has been shown recently, that for 
a linear interferometer sequence 
and arbitrary mixed separable input states the phase 
sensitivity cannot surpass the shot-noise limit \cite{GiovannettiPRL06,PezzePRL09,HyllusArXiv10}
\be
	\Delta\theta_{\rm SN}=\frac{1}{\sqrt N}.
\ee
Hence if a quantum state allows for a phase estimation
scheme with a sub shot-noise (SSN) phase uncertainty, 
it is necessarily entangled. 
We will refer to such states as useful for SSN interferometry or simply as useful. 

\begin{figure}[t]
 \includegraphics[width=0.45\textwidth]{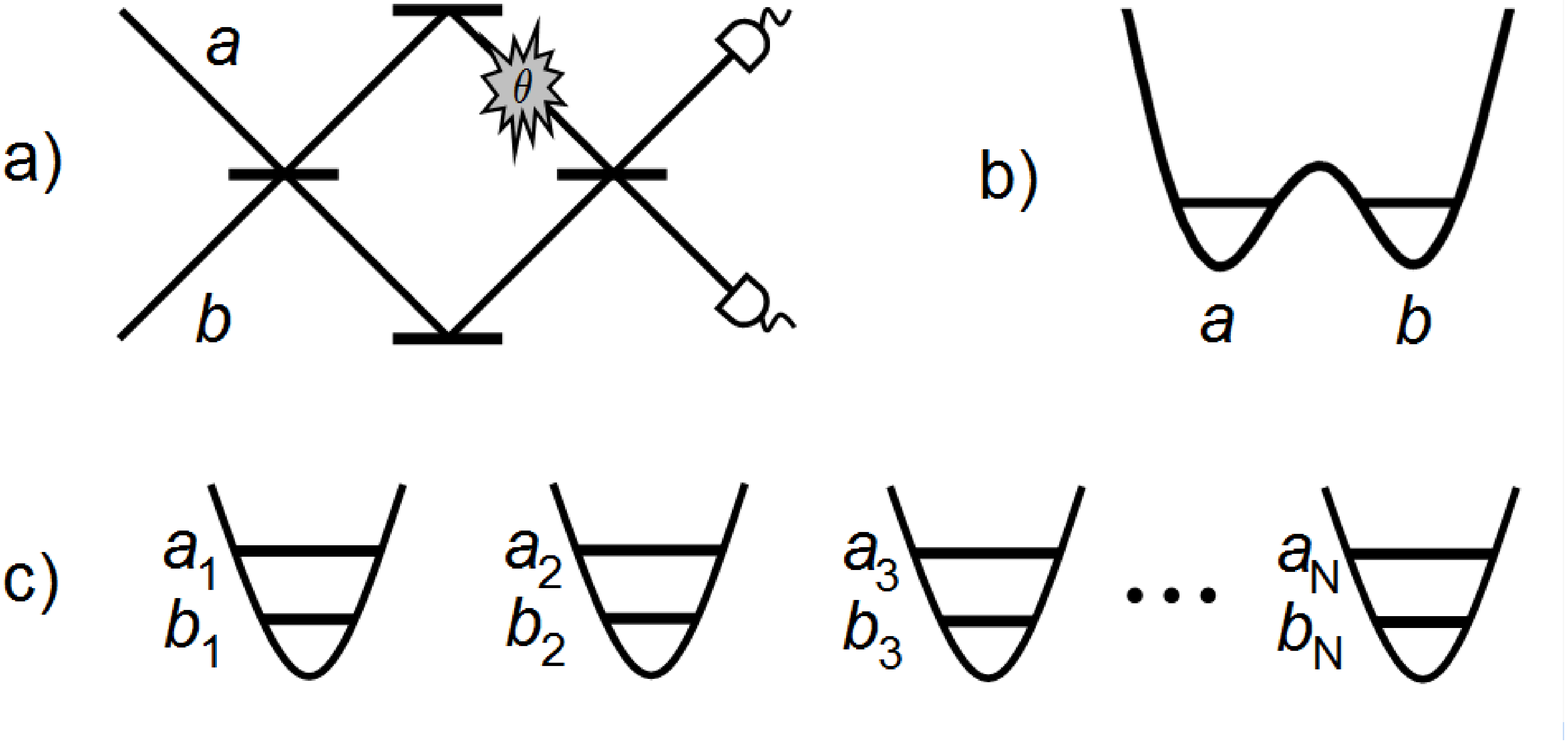}
\caption{Systems that can be used for linear two-state interferometry:
a) archetypical optical Mach-Zehnder interferometer 
as in Refs~\cite{NagataSci07,MatthewsNatPhot09},
b) double-well system as implemented in recent 
experiments on squeezing in BECs 
\cite{EsteveNat08,GrossNat10,RiedelNat10}, and
c) system of single wells as in ion traps 
\cite{LeibfriedNat05,HaeffnerPR08}.
In the first two cases each of the $N$ particles
lives in the subspace of the two states labelled 
by $a$ and $b$, corresponding to momentum states in 
case a) and to the left and the right well in case b).
In case c), there is one particle per well, and
particle $k$ in trap $k$ has the two internal
degrees of freedom $a_k$ and $b_k$ (displayed are 
trap states, while in ion traps typically internal states 
of the ions are used \cite{HaeffnerPR08}).
The interferometer operations acts on the $a$-$b$
subspace in the cases a) and b) and identically
on the subspaces $a_k$-$b_k$ in case c).
In the latter case, the particles are accessible 
individually via the different traps in principle.
They can be treated as distinguishable particles 
labelled by the trap number $k$
if the spacial wavefunctions of 
the particles in the different traps do not overlap 
\cite{PeresBook}.
} \label{fig:MZ}
\end{figure}

In particular, we consider the question whether or not all pure entangled states of a fixed particle number $N$ 
can be made useful if arbitrary local operations can be applied on them before they enter the interferometer. 
We allow for such operations as they correspond to a local change of basis,
and hence cannot create entanglement.
The related problem of finding the local transformation optimizing the interferometric performance of a given input state is of interest for experimental applications, where typically such operations are relatively easy to implement. 
We consider separately the cases where the particles
can or cannot be adressed individually, cf. Fig.~\ref{fig:MZ}
for examples of these situations.

We start by introducing the general framework of parameter estimation with linear interferometers and basic facts about
entanglement in Section~\ref{sec:basics}. 
General observations regarding the local
transformations optimizing the phase sensitivity of
an input state are made in Section~\ref{sec:optimality}. 
The main results concering
the usefulness of pure entangled states under local
transformations are presented Section~\ref{sec:results}. 
Here we also comment on the use of more general
local transformations which are not unitary.
Finally, we apply the results to two important families
of states in Section~\ref{sec:examples}.
We summarize our results in Section~\ref{sec:conclook}.

\section{Basic Concepts}
\label{sec:basics}

\subsection{Linear interferometers and collective operators}

In linear interferometers, such as the Mach-Zehnder interferometer, cf. Fig.~\ref{fig:MZ} a), 
the phase shift is due to the independent 
action of some external effect on each particle. 
We restrict ourselves to the situation that the interferometer is 
performed in a two-level subspace here. The two levels could be two momentum states
as for the Mach-Zehnder interferometer, the two wells of a double-well, or two internal
states of the particles, cf. Fig.~\ref{fig:MZ}.
The corresponding phase transformation 
can be characterized in terms of collective spin operators 
$\hat J_i=\frac{1}{2}\sum_{k=1}^N \hat \sigma_i^{(k)}$, where 
$\hat \sigma_i^{(k)}$ is the $i$-th Pauli matrix acting on particle $k$.
Here and in the following, we label the three Pauli matrices by
$x,y,z$ or by $1,2,3$.
The input state is transformed by $\exp[-i\hat J_{\vec n}\theta]$, where
$\hat J_{\vec n}={\vec n}\cdot \vec{\hat J}$ and $\theta$ is the 
phase shift. For a Mach-Zehnder interferometer consisting of 
a beam splitter $\exp[i\hat J_x\frac{\pi}{2}]$, a phase shift
$\exp[-i\hat J_z\theta]$, and another beam splitter $\exp[-i\hat J_x\frac{\pi}{2}]$,
the effective rotation is  \cite{YurkePRA86}
\be
	U_{\rm MZ}=e^{-i\hat J_x\frac{\pi}{2}}
	e^{-i\hat J_z\theta} e^{i\hat J_x\frac{\pi}{2}}=e^{-i\hat J_y\theta},
\ee
hence ${\vec n}=\hat y$. 
This transformation also describes other applications such as the Fabry-Perot 
interferometer,
Ramsey spectroscopy,
and the Michelson-Morley interferometer.

Note that since the collective spin operators are just sums of single-particle
operators, the transformation factorizes, 
\be
\label{eq:CLU}
e^{-i\hat J_{\vec n}\theta}=e^{-i\hat \sigma_{\vec n}^{(1)}\frac{\theta}{2}}\otimes
e^{-i\hat \sigma_{\vec n}^{(2)}\frac{\theta}{2}}
\otimes\cdots\otimes
e^{-i\hat \sigma_{\vec n}^{(N)}\frac{\theta}{2}},
\ee
where $\hat\sigma_{\vec n}=\vec{\hat\sigma}\cdot{\vec n}$. 
Therefore, this operation acts only locally on the particles, and
no entanglement can be created this way. 
Note that this is true in particular for the beam splitter 
operation $\exp[-i\hat J_x\frac{\pi}{2}]$. 

This is different if a mode picture is used. Let us 
consider the situations a) and b) of Fig.~\ref{fig:MZ}. 
In this case, the beam splitter can turn a separable input 
state $\ket{N_a}\otimes\ket{N_b}$ (written in the Fock basis of 
the two modes $a$ and $b$) into an entangled state,
and the connection of entanglement and SSN interferometry 
is lost. 

We call an operation of the form~(\ref{eq:CLU}) a collective local unitary (CLU) operation, 
since each particle is acted on with the same unitary operator. 
A general local unitary (LU) operation is one which factorizes as well but 
where the unitary operations acting on two different particles 
can be different. Note that if the particles cannot be adressed individually,
as in the cases a) and b) of Fig.~\ref{fig:MZ}, then only CLU operations
can be implemented, while LU operations are available if we can adress
the particles separately, cf. Fig.~\ref{fig:MZ} c).

We will generally work with the particle picture,
and call the eigenstates of the $\hat\sigma_z$ operator 
$\ket{0}$ and $\ket{1}$ such that $\hat\sigma_z\ket{0}=\ket{0}$ 
and $\hat\sigma_z\ket{1}=-\ket{1}$. 
Note that from now on we label the two states by $0$ 
and $1$ instead of $a$ and $b$ as done in 
Fig.~\ref{fig:MZ}.
The eigenstates of the collective spin 
operator $\hat J_z$ will be denoted by $\ket{j,m}$,
where $j=\frac{N}{2}$ 
and $2m$ is the difference of particles in the state $\ket{0}$ and particles 
in the state $\ket{1}$. These states are also known as Dicke-states \cite{DickePR54}. 
They fulfil $\hat{\vec J}^2\ket{j,m}=j(j+1)\ket{j,m}$ and $\hat J_z\ket{j,m}=m\ket{j,m}$.
In general, the eigenvalue $m$ is degenerate. However, the
{\em symmetric} Dicke states $\ket{N/2,m}_S$ are uniquely defined. Here and 
in the following, pure symmetric states are those which are invariant under 
the interchange of any two particles \cite{TothPRL09}. 
Examples for two particles are 
$\ket{1,-1}_S=\ket{1}\otimes\ket{1}\equiv\ket{11}$, $\ket{1,0}_S=(\ket{10}+\ket{01})/\sqrt{2},$
$\ket{1,1}_S=\ket{00}$ and for three particles 
$\ket{3/2,1/2}_S=(\ket{100}+\ket{010}+\ket{100})/\sqrt{3}$.

\subsection{Phase estimation}

\begin{figure}[t]
 \includegraphics[width=0.35\textwidth]{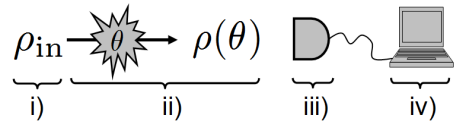}
\caption{A general phase estimation scheme consisting of i) the input state, ii)
the phase transformation, iii) the measurement, and iv) the data processing stage.
} \label{fig:PhaseEstimation}
\end{figure}

In a general phase estimation scenario 
\cite{Helstrom76,Holevo82} (see also \cite{ParisIJQI09} for an introduction),
the initial state $\rho_{\rm in}$ is transformed to $\rho(\theta)$ 
by some transformation depending only on the single parameter $\theta$,
and finally, a measurement is performed. The phase is then estimated
from the results of this measurement. This scheme is schematically depicted
in Fig.~\ref{fig:PhaseEstimation}.

The phase transformation could be, for instance, the operator
$\exp[-i\hat J_y\theta]$ for a Mach-Zehnder interferometer,
as we have seen in the last Section.
A general measurement can be expressed by its positive operator valued measure 
(POVM) elements $\{\hat E(\xi)\}_\xi$ \cite{NielsenBook}.
Depending on the possible outcomes $\xi$, $\theta$ can be estimated from the results 
of these measurements with an estimator $\theta_{\rm est}(\xi)$. 

For so-called unbiased estimators the relation $\bar\theta_{\rm est}=\theta$ holds,
the estimated phase shift is on average equal to the true phase shift. 
The phase sensitivity is defined as the standard deviation of the estimator.
If the estimator is unbiased, it is bounded by the Cram{\'e}r-Rao theorem \cite{Helstrom76,Holevo82} as
\begin{equation}
	\Delta\theta_{\rm est}\ge \frac{1}{\sqrt{m}}\frac{1}{\sqrt{F}}
	\label{eq:CR},
\end{equation}
where $m$ is the number of independent repetitions of the 
measurement and $F$ is the so-called Fisher information.
Fisher's theorem ensures that the bound (\ref{eq:CR}) can be saturated in the 
central limit, typically for large $m$,
with a maximum-likelihood estimator \cite{Cramer46}.

The Fisher information quantifies the statistical distinguishability of 
quantum states along a path described by a single parameter $\theta$
when the measurement $\{\hat E(\xi)\}$ is performed
\cite{WoottersPRD81,BraunsteinPRL94,BraunsteinAP96}. It is defined as
\be
F[\rho(\theta); \{\hat E(\xi)\}]
=\int d\xi P(\xi|\theta)\big[\partial_\theta \log P(\xi|\theta)\big]^2,
\ee
where the conditional probabilities are given by the quantum mechanical
expectation values $P(\xi|\theta)=\tr[\hat E(\xi)\rho(\theta)]$.
This holds for general parameter estimation protocols. In this manuscript,
we only consider estimation protocols for a dimensionless phase
shift and linear interferometers.

The so-called quantum Fisher information $F_Q$ is defined as the 
Fisher information maximized over all possible measurements,
\BE
F_Q[\rho(\theta)]=\max_{\{\hat E(\xi)\}}F\big[\rho(\theta);\{\hat E(\xi)\}\big].  
\EE
For pure input states and for a unitary phase transformation with the generator
$\hat H$, where $\hat H=\hat J_{\vec n}$ for linear two-mode interferometers,
the quantum Fisher information is \cite{BraunsteinPRL94,BraunsteinAP96}
\be
F_Q[\ket{\psi};\hat H]=4\mean{\Delta \hat H^2}_\psi = 4 (\mean{\hat H^2}_\psi- \mean{\hat H}_\psi^2).
\label{eq:FQpure}
\ee
For mixed input states, the quantum Fisher information is given by \cite{BraunsteinPRL94,BraunsteinAP96}
\BE
F_Q[\rho;\hat H]=2\sum_{j,k}(\lambda_j+\lambda_k)\Big(\frac{\lambda_j-\lambda_k}{\lambda_j+\lambda_k}\Big)^2|\bra{j}\hat H\ket{k}|^2,
\label{eq:FQmixed}
\EE
where $\rho=\sum_k \lambda_k \ketbra{k}$ is the spectral decomposition
of the input state and the sum is over terms where $\lambda_j+\lambda_k\neq 0$ only. 

A useful property of $F$, and consequently of $F_Q$, is that it is convex for mixed states, 
{\em i.e.}, if $\rho= p \rho_1 + (1-p)\rho_2$ with $0 \leq p \leq 1$ 
then $F(\rho) \leq p F(\rho_1)+(1-p)F(\rho_2)$ for fixed phase-transformation and
output measurement
\cite{CohenIEEEIT68}, see also Ref.~\cite{FujiwaraPRA01}.

\subsection{Entanglement vs. shot-noise limit}

A pure state of $N$ particles is called fully separable if can be written as a product state,
$\ket{\psi_{\rm fs}}=\otimes_{i=1}^N\ket{\psi^{(i)}}$, where $\ket{\psi^{(i)}}$ is a pure
state of particle $i$. A mixed state is fully separable 
if it can be written as an incoherent  mixture of such product states,
\be\label{eq:fs}
	\rho_{\rm fs}=\sum_k p_k \proj{\psi^{(1)}_k}\otimes\proj{\psi^{(2)}_k}\otimes\cdots\otimes\proj{\psi^{(N)}_k},
\ee
where $\{p_k\}$ is a probability distribution \cite{WernerPRA89}. Any such state
can be generated by local operations and classical communication \cite{WernerPRA89,NielsenBook}.
Non-separable states are entangled, and non-local operations are 
needed for their production.

Recently, it has been shown that for all fully separable input states,
and for any unitary generator of a linear two-mode interferometer
$\hat H=\hat J_{\vec n}$, 
the Fisher information is bounded by the number of particles, 
$F[\rho_{\rm fs};\hat J_{\vec n}]\le N$ \cite{PezzePRL09}.
By the Cram{\'e}r-Rao bound~(\ref{eq:CR}), the phase sensitivity is then
bounded by the shot-noise limit,
\be
	\Delta\theta_{\rm est}\ge \frac{1}{\sqrt{N_{\rm tot}}},
\ee
where $N_{\rm tot}=mN$ is the total number of particles used in the $m$ runs.
Therefore, only entangled input states can reach a sub shot-noise sensitivity.

The so-called Heisenberg limit, {\em i.e.},
the ultimate limit on the phase sensitivity depends on the constraints
on the resources used. If $m$ and $N$ are fixed separately, then the ultimate
sensitivity allowed by quantum mechanics is given by \cite{GiovannettiPRL06}
\be
	\Delta\theta = \frac{1}{\sqrt{m}N}.
\ee
However, the total number of particles used 
in the protocol is $N_{\rm tot}$, and therefore
it is reasonable to consider the bound where
only this number is fixed 
\cite{BraunsteinPRL92b,PezzePRA(R)06}.
The corresponding limit is given by
\be
	\Delta\theta_{\rm HL} = \frac{1}{N_{\rm tot}},
\ee
which can be saturated for $m=1$ only.

We remark that if the interferometer transformation is not
equal to $\exp[-i\hat J_{\vec n}\theta]$, then the shot noise
limit and the Heisenberg limit have to be redefined accordingly. 
Assume, for instance, that the unknown phase shift $\theta$ 
can be applied to a photon a number of times $p$ at will. 
Then a protocol where single photons 
are passing through an interferometer one after the other,
such that photon $k$ experiences the phase shift $p_k\cdot\theta$,
can reach a sensitivity scaling as $\Delta\theta\sim 1/N$
\cite{HigginsNat07,BerryPRA09}.
Here, $N$ is the total number of resources, where not only 
a photon but also the application of the phase shift is counted 
as a resource. 

\subsection{Statement of the problem}

Now we are ready to start the main investigation. 
We want to classify the pure entangled states  with respect
to their usefulness for interferometry. Since entanglement cannot be generated 
by local operations, we allow for such operations to
be applied to the input state. The question we want to answer 
is: can we obtain a bound below the shot-noise limit for every
pure entangled state in this scenario? In other words: 
can $F>N$ be achieved for every pure entangled state?

We consider two cases. (i) If the particles are indistinguishable 
bosons which cannot be individually adressed, 
then the input states have to be symmetric under interchange of the
particles. The results relating separable states to the shot noise
limit is still valid here, but since the state space is reduced,
the only admissible separable states are of the form~(\ref{eq:fs}),
where all single-particle states are identical, 
{\em i.e.}, $\ket{\psi_k^{(i)}}=\ket{\psi_k}$ for all $k$
\cite{Note_indist}. A typical example is the state $\ket{0}^{\otimes N}$
of $N$ particles all entering the Mach-Zehnder interferometer
at the same input port. This situation is present in cases a) and b)
depicted in Fig.~\ref{fig:MZ}.
Only CLU operations can be implemented in this case. 
(ii) The particles (bosons or fermions) can be individually adressed,
for instance because each particle is trapped in a different trap
as in case c) depicted in Fig.~\ref{fig:MZ}.
Then, the particles can be effectively treated
as being distinguishable \cite{PeresBook} and any LU operation can be 
implemented. 

Before we start, let us make two further remarks.
Firstly, optimizing the Fisher information minimizes
the lower bound on the sensitivity~(\ref{eq:CR}).
However, the smallest number $m$ for which this
bound is saturated depends on the input state.
For fixed total resources $N_{\rm tot}=mN$ and
two input states $\ket{\psi}$ and $\ket{\phi}$, it may
may therefore be possible to reach a better
sensitivity $\Delta\theta$ with the state $\ket{\phi}$
even if 
$F(\ket{\psi})>F(\ket{\phi})$ \cite{BraunsteinPRL92b,PezzePRA(R)06}. 
Secondly, the problem we investigate can be viewed 
as one further step in the optimization of the Fisher information
when the phase is generated unitarily:
\be
	F[\rho;\hat H;\{\hat E(\xi)\}_\xi] \le F_Q[\rho;\hat H] \le \max_{U_{\rm L}} F_Q[U_{\rm L}\rho U_{\rm L}^\dagger;\hat H],
\ee
where $U_{\rm L}$ is a local unitary operation.
Both steps preserve the fact that the shot noise limit cannot
be overcome with separable states.

\section{Optimal Fisher information under CLU and LU operations}
\label{sec:optimality}

In this section, we search for the optimal value of the quantum Fisher information
that can be achieved if CLU or LU operations are applied on a pure input state.
We first investigate their effect on $F_Q$ before we find the optimal value of $F_Q$
for CLU operations and an upper bound for LU operations. Finally, we show that
even though we are considering pure states only in this article, similar
results for the optimal values of $F_Q$ hold for mixed states as well.

\subsection{Effect of  CLU and LU operations on $F_Q$}

When the input state $\ket{\psi_{\rm in}}$ is transformed by a local unitary transformation
$U_{\rm L}=U_1\otimes U_2\otimes \cdots \otimes U_N$, 
then the quantum Fisher information Eq.~(\ref{eq:FQpure}) changes as 
\be
	F_Q'=4\mean{\Delta \hat H^2}_{U_{\rm L}\psi_{\rm in}}=4\mean{\Delta \hat H'^2}_{\psi_{\rm in}},
\ee
where $\hat H'=U_{\rm L}^\dagger \hat H U_{\rm L}$. Hence, for $\hat H=\hat J_y$ as in the Mach-Zehnder
interferometer, applying a LU operation to the initial state is equivalent
to a local transformation of the interferometer operation according to
$\hat J'=\frac{1}{2}\sum_{k=1}^N U_k^\dagger \hat\sigma_y U_k$.
The relation $U^\dagger\hat{\vec\sigma} U=O\hat{\vec\sigma}$ holds,
where $O$ is an orthogonal matrix, hence
a unitary transformation of the vector of Pauli 
matrices corresponds to a rotation \cite{SakuraiBook}.
It follows that
\be
	\hat J'=\frac{1}{2}\sum_{k=1}^N {\vec n}^{(k)} \cdot \hat{\vec\sigma}.
	\label{eq:Jprime}
\ee 
Here
${\vec n}^{(k)}=O_k^T {\hat y}$, and ${\hat y}$ is the unit vector pointing in
the $y$-direction. 

Therefore, changing the input state with a LU operation is equivalent to 
a change of the local directions of the spins. A collective spin operator
is in general acting differently on the spins after this operation. 
If a CLU operation is applied, where $U_k=U$ for all $k$,
then the collective operator remains collective, 
only its direction is changed. 

\subsection{Optimum under CLU operations}
\label{sec:opt_CLU}

Given a general pure state $\ket{\psi}$, the optimal direction 
${\vec n}_{\rm max}$ of the generator $\hat J_{\vec n}$ and the maximal 
$F_Q$ can be determined directly.\\[1mm]
\noindent 
{\bf Observation 1}. The maximal $F_Q$ that can be achieved for 
$\hat H=\hat J_{\vec n}$ when ${\vec n}$ can be optimized over
is given by $4\lambda_{\rm max}[\gamma_C]$, where 
$\lambda_{\max}$ is the maximal eigenvalue of the real 
$3\times 3$ covariance matrix $\gamma_C$ with entries
\be
[\gamma_C]_{ij}=\frac{1}{2}\mean{\hat J_i \hat J_j+\hat J_j \hat J_i}-\mean{\hat J_i}\mean{\hat J_j},
\ee
and the optimal direction ${\vec n}_{\rm max}$ is the corresponding eigenvector
\cite{Note_Obs1}.
We will also call  $\gamma_C$ the collective covariance matrix.
\\[1mm]
\noindent 
{\em Proof}. For $\hat H={\vec n}\cdot \vec J$ we have 
$F_Q=4\mean{(\Delta \hat J_{\vec n})^2}=4{\vec n}^T \gamma_C{\vec n}$
since ${\vec n}$ is real. It is known from linear algebra that this 
expression is maximized by choosing 
${\vec n}={\vec n}_{\rm max}$ as the eigenvector corresponding to
the maximal eigenvalue.
\proofend

The matrix $\gamma_C$ has appeared before in the context of interferometry 
\cite{RivasPRA08}
and in the derivation of the optimal spin squeezing inequalities for entanglement 
detection \cite{TKGB09}. 
The results presented in the latter article allow
for a different proof of the fact that separable states 
cannot beat the shot noise limit 
\cite{Note_Obs1b}.

\begin{figure}[t]
 \includegraphics[width=0.34\textwidth]{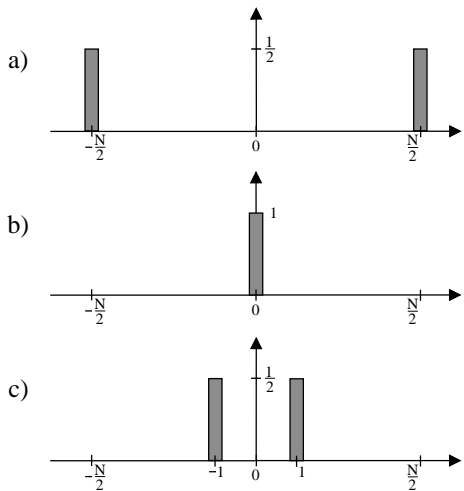}
\caption{Examples of symmetric states that we apply Observation 1 on given in the 
basis of symmetric Dicke states. 
a) NOON state, b) twin-Fock state, 
c) state considered in Ref.~\cite{PezzePRA(R)06}. 
Plotted are the squared absolute values of the weights of the 
symmetric Dicke states in the superpositions.} 
\label{fig:symexamples}
\end{figure}

Let us consider as examples of the usefulness of Observation 1 
three prominent symmetric states which are known to provide SSN sensitivity.
Their weights in the basis of symmetric Dicke states 
$\ket{\frac{N}{2},m}_S$ are depicted in 
Fig.~\ref{fig:symexamples}.

a) The so-called NOON state is given by
\cite{BollingerPRA(R)96,LeeJMO02}
\be
\ket{{\rm NOON}}=\frac{1}{\sqrt{2}}\Big(\ket{0}^{\otimes N}+\ket{1}^{\otimes N}\Big).
\ee
For this state, we find $4\gamma_C^{\rm NOON}=\diag(N,N,N^2)$. 
The NOON state achieves the maximal value of the Fisher information $F_Q=N^2$ \cite{GiovannettiPRL06}
when the generator of the phase shift is $\hat J_z$, while it gives sensitivity at the 
shot-noise limit if a collective operator in the $x-y$ plane is chosen 
instead. Hence if a NOON state is entering a normal Mach-Zehnder interferometer,
it has to be rotated first by $\exp(\pm i\frac{\pi}{2}\hat J_x)$ in order
to reach the optimal sensitivity. This happens because only the $\hat J_z$ 
operator leads to the maximal relative phase shift $\exp[-i N\phi]$ 
between the two states in the superposition of the NOON state.
Similar corrections have to be applied in the Ramsey scheme 
originally considered in Ref.~\cite{BollingerPRA(R)96}.

b) Another state promising SSN sensitivity 
is the Twin-Fock state \cite{HollandPRL93}
\be
	\ket{{\rm TF}}=\ket{\frac{N}{2},0}_S.
\ee
In this state, half of the particles are in the state $\ket{0}$ and
the other half is in the state $\ket{1}$. Note that this state is a product
state in a mode picture, $\ket{N/2}_0\otimes\ket{N/2}_1$, while
it is multipartite entangled in the particle picture we use.
We find $4\gamma_C^{\rm TF}=(\frac{N^2}{2}+N)\diag(1,1,0)$,
hence the SSN sensitivity for interferometry with the generator 
$\hat J_{\vec n}$ is bounded identically by the Cram{\'e}r-Rao lower
bound for any ${\vec n}$ in the $x-y$ plane, while the state is
insensitive to the phase change if ${\vec n}=\hat z$.

c) The third example is the following state of an even number of particles,
which offers advantages when used in a Mach-Zehnder interferometer
with a Bayesean estimation protocol \cite{PezzePRA(R)06}
\be
	\ket{{\rm PS}}=\frac{1}{\sqrt{2}}\Big(\ket{\frac{N}{2},1}_S
	+\ket{\frac{N}{2},-1}_S\Big),
\ee
which yields 
$4\gamma_C^{\rm PS}=\diag(\frac{3}{4}N^2+\frac{3}{2}N-2,\frac{1}{4}N^2+\frac{1}{2}N-2,4)$.
We expressed the states with the symmetric Dicke states $\ket{j,m}_S$
introduced above.
While both for ${\vec n}=\hat y$ and ${\vec n}=\hat x$ the quantum Fisher information
is larger than $N$, it is largest for ${\vec n}=\hat x$.

\subsection{Optimum under LU operations}

When general LU operations are applied, then the quantum Fisher information 
takes the form
$F_Q[\ket{\psi};\hat J']=4\mean{(\Delta J')^2}={\vec m}^T \gamma_R {\vec m}$,
where we introduced the real $3N \times 3N$ covariance 
matrix $\gamma_R$ with entries
\begin{align}\label{gammaR}
[\gamma_R]_{(k_1,i_1)(k_2,i_2)}
&
=
\frac{1}{2}
(
\mean{\hat\sigma^{(k_1)}_{i_1}	\hat \sigma^{(k_2)}_{i_2}} 
+
\mean{\hat\sigma^{(k_2)}_{i_2}	\hat\sigma^{(k_1)}_{i_1}} 
)
\nonumber
\\
&
- \mean{\hat\sigma^{(k_1)}_{i_1}}\mean{\hat\sigma^{(k_2)}_{i_2}},
\end{align}
and the real vector
${\vec m}^T=([{\vec n}^{(1)}]^T,[{\vec n}^{(2)}]^T,\ldots,[{\vec n}^{(N)}]^T)$.
The matrix entries are parametrized by two double indices $(k_1,i_1)$ and $(k_2,i_2)$.
The optimal value of $F_Q$ is hence given by the solution of the problem
\be
F_Q^{\rm max}=
\max_{\vec m}  {\vec m}^T \gamma_R {\vec m}\Big|_{[{\vec n}^{(k)}]^T {\vec n}^{(k)} =1\ \forall k}.
\label{eq:nonsymproblem}
\ee
A simple upper bound on $F_Q^{\rm max}$ 
can be obtained by relaxing the $N$ constraints $[{\vec n}^{(k)}]^T {\vec n}^{(k)}=1$
to the single constraint ${\vec m}^T {\vec m}=N$:
\\[1mm]
\noindent 
{\bf Observation 2}. 
The maximal $F_Q$ for $\hat H=\hat J_y$ 
that can be achieved when arbitrary LU operations 
can be applied on the input state is bounded by
\be
F_Q^{\rm max}\le\max_{\vec m} {\vec m}^T \gamma_R m\big|_{{\vec m}^T {\vec m}=N} 
=N\lambda_{\rm max}[\gamma_R].
\label{eq:UB}
\ee
Equality holds in relation~(\ref{eq:UB}) iff there is a vector ${\vec m}^*$
optimizing problem~(\ref{eq:nonsymproblem}) which is an eigenvector of 
$\gamma_R$ corresponding to the maximal eigenvalue. 
If ${\vec m}^*$ is the eigenvector corresponding to the maximal
eigenvalue of $\gamma_R$ which fulfills all the $N$ constraints
$[{\vec n}^{(k)}]^T {\vec n}^{(k)}=1$, then 
local directions can be converted into LU transformations
as in Observation 1 \cite{Note_Obs1}. 

This simple Observation, which can be proven in the same way as 
Observation 1, will turn out to be very useful for the proof 
Proposition 4 below and also for the examples discussed in Section~\ref{sec:examples}.

Here we obtained an upper bound on the maximal $F_Q$ that can
be obtained when arbitrary LU operations are available
by allowing to optimize over more general operations.
In turn, a simple {\em lower} bound can be obtained from
Observation 1, since in this case the operations are more restricted.

We remark that the covariance matrix $\gamma_R$ with entries 
given in Eq.~(\ref{gammaR}) has appeared previously in studies 
of macroscopic entanglement \cite{ShimizuPRL02,MorimaePRA05}.
Given a pure state $\ket{\psi}$, then the index $p\in[1,2]$
introduced in these references indicates the presence
of macroscopic entanglement if $p=2$. Similar to our case,
its computation involves the maximization of the variance
of a local operator $\hat A=\sum_{k=1}^N \alpha_k\vec n^{(k)}\cdot\vec{\hat\sigma}$, 
where the $\alpha_k$ fulfill $\sum_{k=1}^N |\alpha_k|^2=N$.
Due to the additional parameters $\{\alpha_k\}_k$,
the problem is then of the form 
$\max_{\vec m} \vec m^\dagger \gamma_R \vec m\big|_{\vec m^\dagger \vec m=N}$,
and the maximum can always be reached by the
maximal eigenvector and the corresponding eigenvalue 
\cite{MorimaePRA05}. Here the dagger $\dagger$ appears 
instead of the transposition $T$ since the $\alpha_k$ are not 
restricted to be real \cite{MorimaePRA05}.

\subsection{Optimum for mixed states}
\label{sec:mixedstates}

Even though in this article we are only considering pure states, we would
like to note that Observation 1 also holds in the case of mixed input states
$\rho=\sum_k \lambda_k \proj{k}$
when the collective covariance matrix  $\gamma_C$ is replaced by 
the matrix $\Gamma_C$ with entries
\be
[\Gamma_C]_{ij}=\frac{1}{2}\sum_{l,m}(\lambda_l+\lambda_m)\Big(\frac{\lambda_l-\lambda_m}{\lambda_l+\lambda_m}\Big)^2 
	\bra{l}\hat J_{i}\ket{m}\bra{m}\hat J_{j}\ket{l}.
\ee
In analogy, Observation 2 holds if the matrix $\gamma_R$ is 
replaced by the matrix $\Gamma_R$ with entries
\bea
  &&[\Gamma_R]_{(k_1,i_1),(k_2,i_2)}=\\
	&&\frac{1}{2}\sum_{l,m}(\lambda_l+\lambda_m)\Big(\frac{\lambda_l-\lambda_m}{\lambda_l+\lambda_m}\Big)^2
  \bra{l}\hat \sigma_{i_1}^{(k_1)}\ket{m}\bra{m}\hat \sigma_{i_2}^{(k_2)}\ket{l}.\nonumber
\eea
This follows directly from the form of $F_Q$ for mixed states when the 
phase comes from unitary evolution generated by $\hat J_{\vec n}$
and $\hat J'$, respectively, see Eq.~(\ref{eq:FQmixed}).
Note that the matrices $\Gamma_C$ and $\Gamma_R$ are symmetric because of the sums over $l$ 
and $m$. For pure states, they reduce to $\gamma_C$ and $\gamma_R$, respectively.

\section{Usefulness of pure entangled states}
\label{sec:results}

Now we are prepared to consider the general question: are
all pure entangled states useful for SSN interferometry
under CLU and LU operations? In the first part of this section,
we will consider pure symmetric states and CLU operations,
corresponding to the situation in a system of $N$ bosons 
which cannot be individually
adressed, as in the cases a) and b) depicted in Fig.~\ref{fig:MZ}. 
If the input state is symmetric but general LU operations
can be applied, we find that $F_Q$ cannot be increased beyond the value
obtained with the optimal CLU operation.
The results allow us to draw conclusions on the usefulness 
of general states. In the final part of this section, we will briefly
comment on how these results change when more general local operations
than CLU and LU are available.

\subsection{Reduced states of pure symmetric states}
\label{sec:redsym}

We start by considering reduced density matrices of pure symmetric states.
This will be very useful for the proofs presented later. The reduced density 
matrix for two particles of any state $\ket{\psi}$ can always
be written as
\be
	\rho^{(r)}=\frac{1}{4}\sum_{i,j=0}^{3}\lambda_{ij}\hat\sigma_i\otimes\hat\sigma_j,
	\label{eq:rhored}
\ee
where $\lambda_{ij}=\mean{\hat\sigma_i\otimes\hat\sigma_j}_{\psi}$,
$\hat\sigma_0=\Eins$, and $\hat\sigma_{1,2,3}=\hat\sigma_{x,y,z}$. Normalization is
ensured by $\lambda_{00}=1$.

If $\ket{\psi}$ is symmetric under the interchange of particles,
then the matrix $\lambda$ is not only real, but also symmetric,
and the diagonal elements fulfil
\cite{TothPRL09}
\be
	\sum_{i=1}^3 \lambda_{ii}=1.
	\label{eq:symsum}
\ee
Note that this holds for the case $N=2$, where $\rho^{(r)}=\proj{\psi}$,
and also for the case $N>2$, since then the reduced density matrix also acts on 
the symmetric subspace only.

If we consider CLU transformations of $\ket{\psi_S}$, then 
$\rho^{(r)} \to U\otimes U\ \rho^{(r)}\  U^\dagger\otimes U^\dagger$.
Since $U\hat{\vec \sigma}U^\dagger = O^T \hat{\vec \sigma}$ as mentioned before,
$\lambda$ transforms as
\be
	\lambda\equiv 
	\left(\begin{array}{cc}
			1 & {\vec s}^T\\
			{\vec s} & T
		\end{array}
	\right)
	\to 
	\left(\begin{array}{cc}
			1 & {\vec s}^T O\\
			O^T{\vec s} & \bar T
		\end{array}
	\right)\equiv \bar\lambda,
	\label{eq:lambda}
\ee
where ${\vec s}$ is a column vector with entries $s_i=\mean{\hat\sigma_i}$, 
$T$ a symmetric $3\times 3$ matrix with entries $T_{ij}=\mean{\hat\sigma_i\otimes \hat\sigma_j}$
for $i,j=1,2,3$, and $\bar T=O^T T O$.
The condition (\ref{eq:symsum}) corresponds to $\tr[T]=1$.
Since $-1\le \lambda_{ij}\le 1$ holds, only one of the diagonal 
elements $T_{ii}$ can be negative. Further, if one element is negative,
then the other two diagonal elements have to be strictly positive.

\subsection{CLU operations}

Here, we consider pure symmetric entangled states under CLU operations. This is realized in a bosonic system where all
particles can be in two external states, 
for instance.
In this situation the states can be completely characterized with respect to their usefulness, 
and it turns out that any symmetric state is 
useful, apart from superpositions of $\ket{0}^{\otimes N}$
and $\ket{1}^{\otimes N}$ with significantly different weights. 
We directly state the result and present the proof afterwards.
\\[1mm]
\noindent
{\bf Proposition 3}. For a pure, symmetric, and entangled state $\ket{\psi_S}$
there is a direction ${\vec n}$ such that $F_Q(\ket{\psi_S},\hat J_{\vec n})>N$
except for the following family of states of $N>2$ qubits:
\be 
	\ket{\psi_S}=\sqrt{q}\ket{0}^{\otimes N}+e^{i\phi}\sqrt{1-q}\ket{1}^{\otimes N}
	\label{eq:ghz_p}
\ee 
up to a CLU operation and\be
	q\le \frac{1}{2}\Big(1-\sqrt{\frac{N-1}{N}}\Big)
	\quad 
	\textrm{or}
	\quad
	q\ge\frac{1}{2}\Big(1+\sqrt{\frac{N-1}{N}}\Big).
	\label{eq:pcond}
\ee
\\[1mm]
\noindent
{\em Proof}. The form of $F_Q(\ket{\psi},\hat J_{\vec n})$ is
\bean
	&& 4\mean{(\Delta\hat J_{\vec n})^2}=\mean{\sum_{k,l} \hat\sigma_{\vec n}^{(k)} \hat\sigma_{\vec n}^{(l)} }-\mean{\sum_k \hat\sigma_{\vec n}^{(k)} }^2\\
	&=&N-\sum_k\mean{\hat\sigma_{\vec n}^{(k)}}^2+2\sum_{k<l}\mean{\hat\sigma_{\vec n}^{(k)}
	 \hat\sigma_{\vec n}^{(l)}}-\mean{\hat\sigma_{\vec n}^{(k)} }\mean{\hat\sigma_{\vec n}^{(l)} }.
\eean
For symmetric states, the terms $\mean{\hat\sigma_{\vec n}^{(k)}}$ and $\mean{\hat\sigma_{\vec n}^{(k)} \hat\sigma_{\vec n}^{(l)}}$ 
do not depend on the sites $k$ and $l$, and hence
\bean
	4\mean{(\Delta\hat J_{\vec n})^2}
	&=&N(1-\mean{\hat\sigma_{\vec n}}^2)+N(N-1)(\mean{\hat\sigma_{\vec n} \hat\sigma_{\vec n} }-\mean{\hat\sigma_{\vec n} }^2)\\
	&=&N+N(N-1)\mean{\hat\sigma_{\vec n} \hat\sigma_{\vec n} }-N^2\mean{\hat\sigma_{\vec n}}^2
\eean
where we left out the particle indices.
It follows that for pure symmetric states
\be
 F_Q[\psi_S;\hat J_{\vec n}]>N \quad
 \Leftrightarrow\quad	\mean{\hat\sigma_{\vec n} \hat\sigma_{\vec n} } > \frac{N}{N-1}\mean{\hat\sigma_{\vec n}}^2.
 \label{eq:symcond}
\ee
Hence the task is to see whether or not it is possible for any pure symmetric 
state to find a CLU operation or a direction ${\vec n}$
such that condition (\ref{eq:symcond}) holds.

We can choose a CLU operation such that $\mean{\hat\sigma_x}=\mean{\hat\sigma_y}=0$,
$O^T {\vec s}=(0, 0, \delta)^T$. Now we have to consider several cases:
(i) Let us assume that the elements $\bar T_{ij},\ i,j=1,2$ are not 
all equal to zero. 
Since $\mean{\hat\sigma_x}=\mean{\hat\sigma_y}=0$, then
if we can make $\mean{\hat\sigma_x\hat\sigma_x}$ or $\mean{\hat\sigma_y\hat\sigma_y}$ positive 
then condition (\ref{eq:symcond}) is fulfilled for the respective
direction. 
If there are non-zero elements $\bar T_{ij},\ i,j=1,2$, 
then the trace of this submatrix might be zero, 
but the eigenvalues will be different from zero.
Then they have to be both 
different from zero, and only one of them can be negative. 
Hence there is an orthogonal transformation $1\oplus \tilde O\oplus 1$ which
makes $\lambda_{11}$ 
positive while keeping 
${\vec s}=(0,0,\delta)^T$. Therefore, we can make $\mean{\hat\sigma_x\hat\sigma_x}$
positive and fulfil
condition (\ref{eq:symcond}) for ${\vec n}=\hat x$.

(ii) If all elements $\bar T_{ij}$ are equal to zero for $i,j=1,2$,
then $\bar T_{33}=1$ due to Eq.~(\ref{eq:symsum}). What kind of states 
$\ket{\psi_S}$ are compatible with these values? Only those of 
the form of Eq.~(\ref{eq:ghz_p}). This can be seen as follows:
we can expand $\ket{\psi_S}=\sum_{m=-N/2}^{N/2} c_m \ket{N/2,m}_S$
in the basis of symmetric Dicke states.
Then 
$1=\mean{\hat\sigma_z \hat\sigma_z}=\sum_m |c_m|^2\bra{N/2,m}\hat\sigma_z \hat\sigma_z\ket{N/2,m}
=\sum_m |c_m|^2$, where the latter equality comes from the normalization of $\ket{\psi_S}$. 
It follows that $c_m$ can only be different from zero
if $\bra{N/2,m}\hat\sigma_z \hat\sigma_z\ket{N/2,m}=1$, which is the case 
for $m=\pm N/2$ only. These are the states of Eq.~(\ref{eq:ghz_p})
with the notation
$c_{N/2}=\sqrt{q}$ and $c_{-N/2}=e^{i\phi}\sqrt{1-q}$.

For $N>2$, the coefficients $s_i$ and $T_{ij}$ ($i,j=1,2$)
vanish for any value of $q$.
In this case, $\mean{\sigma_z}=2(q-\frac{1}{2})$, and
condition~(\ref{eq:symcond}) reads $(q-\frac{1}{2})^2<\frac{(N-1)}{4N}$.
This condition is violated if Eq.~(\ref{eq:pcond}) holds.
This suggests that if $q$ is too close to $0$ or to $1$, 
then $F_Q\le N$. What is left to show is that 
there is no other direction ${\vec n}$ where $F_Q>N$ for this state. 
This follows directly from Observation 1 since 
$\gamma_C=\diag(\frac{N}{4},\frac{N}{4},\frac{N^2}{4}[1-(2q-1)^2])$
is diagonal already. 
Hence there is no better basis, and if Eq.~(\ref{eq:pcond})
holds, then the entanglement of the 
state (\ref{eq:ghz_p}) is not useful for 
SSN interferometry in any direction ${\vec n}$. 

In contrast, for $N=2$, the coefficients 
$s_i$ and $T_{ij}$ for $i,j=1,2$ vanish 
only if $q=0$ or $q=1$, {\em i.e.}, if $\ket{\psi_S}$ is a product state. 
It follows that any pure symmetric entangled 2-qubit state
is useful for SSN interferometry. 

\proofend

We would like to point out three things concerning the states~(\ref{eq:ghz_p}).
(i) the region where the states are not useful shrinks with 
increasing $N$. (ii) when $q$ is changed such that 
the states change from being useful to not being useful,
then the optimal direction ${\vec n}$ changes from $\hat z$ 
to any direction in the $x-y$ plane. This is not surprising,
since for the product states $\ket{0}^{\otimes N}$ and 
$\ket{1}^{\otimes N}$, the variance of $\hat J_{\vec n}$ 
is maximized for ${\vec n}$ lying in the $x-y$ plane, 
while the variance of the NOON state 
$\frac{1}{\sqrt{2}}(\ket{0}^{\otimes N}+\ket{1}^{\otimes N})$
is maximized for ${\vec n}=\hat z$ as seen in section~\ref{sec:opt_CLU}.
However, one could have
expected a smooth transition from ${\vec n}=\hat z$ to
the $x-y$ plane. (iii) the states of Eq.~(\ref{eq:ghz_p})
are not separable with respect to any partition if $q\neq 0$ and $q\neq 1$,
and hence genuinely multipartite entangled, but still 
of no use for SSN interferometry when condition~(\ref{eq:pcond}) holds
and only CLU can be applied to the input state.

\subsection{LU operations}

We have found that states of the form~(\ref{eq:ghz_p}) are not useful
for sub shot-noise interferometry if the condition~(\ref{eq:pcond}) holds
and if only CLU operations can be applied. 
It is natural to ask whether or not this can be changed by applying 
{\it arbitrary} local unitary operations on this
state. 
It turns out, however, that this more general class of 
transformations does not help. This is the content of Proposition 4 below.
Hence not all pure entangled states are useful for SSN interferometry,
even if arbitrary LU operations can be applied to the input state.
The main results of this article regarding this question are summarized in Theorem 5.
\\[1mm]
\noindent
{\bf Proposition 4}. For a pure, symmetric, and entangled state $\ket{\psi_S}$ 
under LU operations the maximum quantum Fisher information is obtained
by choosing a collective spin vector with ${\vec n}_{\rm max}$ determined
as stated in Observation 1. For $N>2$, any non-collective operation
leads to a strictly smaller value of $F_Q$.
\\[1mm]
\noindent
{\em Proof}. In order to apply Observation 2, we first have to construct $\gamma_R$
as defined in Eq.~(\ref{gammaR}).
The terms $\mean{\hat\sigma_i^{(k)}}$ and 
$\mean{\hat\sigma_i^{(k)} \hat\sigma_j^{(l)}}$ 
do not depend on the sites $k$ and $l$ if $\ket{\psi_S}$ is symmetric. The resulting
covariance matrix has the block-form
\be
	\gamma_R=\left(
	\begin{array}{ccccc}
		A & B & B & ... & B\\
		B & A & B & ... & B\\
		\vdots & \vdots & \vdots & \ddots & \vdots\\
		B & B & B & ... & A		
	\end{array}
	\right),
	\label{eq:gammaS}
\ee
where $A_{ij}=\delta_{ij}-\mean{\hat\sigma_i}\mean{\hat\sigma_j}$ 
and $B_{ij}=\mean{\hat\sigma_i\hat\sigma_j}-\mean{\hat\sigma_i}\mean{\hat\sigma_j}=B_{ji}$ are $3\times 3$ matrices.
With the notation introduced above Eq.~(\ref{eq:lambda}), we can write 
$A=\Eins-{\vec s}{\vec s}^T$ and $B=T-{\vec s}{\vec s}^T$. 
The rank of $\gamma_R$ is in general full, but there are
at most $6$ distinct eigenvalues. This can be seen as follows:
If we find the three eigenvectors ${\vec a}_k$ of the matrix
$[A+(N-1)B],$ then we can directly construct three eigenvectors of $\gamma_R$ 
which are fully symmetric under interchange of the blocks, 
namely ${\vec x}_k^T=({\vec a}_k^T,{\vec a}_k^T,\ldots,{\vec a}_k^T)$, 
where $k=1,2,3$.
Furthermore, if we find three eigenvectors ${\vec b}_k$ of
the matrix $[A-B]$, we obtain $3(N-1)$ linearly
independent eigenvectors of $\gamma_R$ of the form 
$[{\vec y}^{(j)}_k]^T=({\vec b}_k^T,0,...,0,-{\vec b}_k^T,0,...,0)$,
where the second vector $-{\vec b}_k$ is located at the positions of 
block $j$, $j=2,3,...,N$. 
These vectors are orthogonal to the vectors ${\vec x}_k$ by construction, 
so the spectrum of $\gamma_R$ is given by the eigenvalues of the matrices 
$[A+(N-1)B]$ and $[A-B]$. Let us denote by $\lambda_1$ the largest eigenvalue 
of the first matrix and by $\lambda_2$ the largest eigenvalue of the second matrix.
If $\lambda_1\ge\lambda_2$, then the optimal $F_Q$ can be reached by a 
collective spin operator with all spin operators pointing in the same direction,
while if the inequality holds strictly, $\lambda_1>\lambda_2$, 
then it is clear that the optimal $F_Q$ reached with 
a collective operator is strictly larger than the largest $F_Q$ that 
can be achieved with a non-collective spin operator.

Comparing $\lambda_1$ and $\lambda_2$ is equivalent to comparing the 
eigenvalues of $(N-1)T-N{\vec s}{\vec s}^T$ and $-T$.
Let us denote the eigenvalues of the matrix $T$ by $t_i$, $i=1,2,3$ 
and order them increasingly. Due to Eq.~(\ref{eq:symsum}), they fulfil 
$t_1+t_2+t_3=1$. The largest eigenvalue of $-T$ is hence given by $-t_1$. 
We have to consider several cases:
(i) if $t_1 > 0$, then $-T$ has no positive eigenvalues, 
whereas $(N-1)T$ has only positive eigenvalues. 
Hence there is always a vector ${\vec r}_\perp$ orthogonal to ${\vec s}$
such that ${\vec r}_\perp^T[(N-1)T-N{\vec s}{\vec s}^T]{\vec r}_\perp > 0 > -t_1$,
which implies $\lambda_1>\lambda_2$.
(ii) If $t_1\le 0$, $-T$ has at most one positive eigenvalue $|t_1|$.
There is a two-dimensional subspace $\tilde S$
such that for ${\vec r}\in\tilde S$, ${\vec r}^T T {\vec r}\ge |t_1|$ holds
since $t_3\ge t_2\ge|t_1|$. 
The last inequality holds since $|t_i|\le 1$ for all $i$. 
Hence there is a vector ${\vec r}_\perp\in\tilde S$
orthogonal to ${\vec s}$ with 
${\vec r}_\perp^T[(N-1)T-N{\vec s}{\vec s}^T]{\vec r}_\perp \ge (N-1)t_2 \ge |t_1|$,
implying $\lambda_1\ge \lambda_2$.

So far we have shown that we can always choose a symmetric collective 
operator $\hat J_{\vec n}$. Let us focus now on the cases where $\lambda_1=\lambda_2$
holds, where also non-symmetric collective operators may reach the optimal $F_Q$.
This may happen if $N=2$ and $t_2=|t_1|$ or if $N>2$ and $t_1=t_2=0$.
In the first case, symmetric vectors $({\vec n}^T, {\vec n}^T)^T$ and
antisymmetric vectors $({\vec n}^T, -{\vec n}^T)^T$ always reach the same 
optimum unless $t_1=t_2=0$, when the state is separable.
This can be seen by a direct calculation of $\lambda$ 
with the general symmetric state
$\ket{\psi_S}=c_1\ket{1,1}+c_{1,-1}\ket{1,-1}+c_0\ket{1,0}$ and by requiring
that $T$ is diagonal with $t_1=-t_2$.
In the second case, $T=\diag[0,0,1]$. As 
mentioned in the proof of Proposition 3, this is only possible for states
of the form~(\ref{eq:ghz_p}), for which ${\vec s}^T=(0,0,\delta)$, 
where $\delta= 2(q-\frac{1}{2})$.
Then the condition that $(N-1)T-N{\vec s}{\vec s}^T=\diag[0,0,(N-1)-N\delta^2]$
has a strictly larger eigenvalue than $|t_1|=0$ is fulfilled
unless condition~(\ref{eq:pcond}) 
holds. If it holds then $F_Q=N$, and this can be reached by
choosing ${\vec n}^{(k)}=(c_1^{(k)},c_2^{(k)},0)^T$ for any $c_{1,2}^{(k)}$,
as can be seen directly by writing down $\gamma_R$ from Eq.~(\ref{eq:gammaS}) 
in this case.
\proofend

Summarizing, this allows us to formulate a central result of this article.\\[1mm]
\noindent
{\bf Theorem 5}. Allowing for general LU operations to be applied
on the input state, then for $N=2$, any pure entangled state is useful for 
SSN interferometry. For $N>2$, there are pure entangled states
which are not useful even if they can be transformed by arbitrary 
LU transformations. The pure entangled symmetric states which are not 
useful are completely characterized by Proposition 3.
\\[1mm]
\noindent
{\em Proof.} Proposition 4 implies that even allowing for any LU
operation does not make the states~(\ref{eq:ghz_p}) useful for 
SSN interferometry if condition~(\ref{eq:pcond}) holds. Therefore, for $N>2$, 
there are pure entangled states which cannot be made useful.
For $N=2$ we have seen already that all states of the form 
$\sqrt{q}\ket{00}+e^{i\phi}\sqrt{1-q}\ket{11}$ are useful 
unless $q=0$ or $q=1$. In this case any state
can be brought into this form by a local change of basis,
therefore, any pure entangled state of two qubits is useful
for sub shot-noise interferometry.  \proofend

The result that all entangled states with $N=2$ particles 
lead to $F_Q>N$ for some change of the local basis also follows directly
from results obtained for the Wigner-Yanase skew information $I(\rho,\hat H)$ 
depending on a state $\rho$ and an observable $\hat H$ \cite{WignerPNAS63}.
For any pure entangled state $\ket{\psi_{\rm ent}}$ of $N=2$ particles, 
it has been shown that $4I(\ket{\psi_{\rm ent}},\hat H)>2$ can be achieved
by local rotations \cite{ChenPRA05}. This proof carries over to the Fisher
information since for pure states, the quantities are related by 
$F(\ket{\psi},\hat H)=4I(\ket{\psi},\hat H)$.

\subsection{More general local operations}

So far we considered the scenario that a single copy of a pure state is 
used to perform a phase estimation protocol. We allowed for local manipulations
of this state prior to the experiment. 
Typically, investigations of quantum entanglement assume that the
parties controlling the particles are very far apart 
and that they can only perform local operations on their system
and classical communication (LOCC) \cite{NielsenBook}. 
In this scenario, the particles can be treated effectively as
distinguishable \cite{PeresBook} and general LU can be applied.
More general local measurements can be performed when each party is 
allowed to add local particles, so-called ancillas, and to perform LU
operations and measurements on the ancilla particles, discarding them
after the operation \cite{NielsenBook}. They can be described
with so-called Kraus operators $\hat A_i$, where $i$ labels the results
of the local measurements. They fulfil $\sum_i \hat A_i^\dagger \hat A_i=\Eins$ 
and transform the initial state as 
$\ket{\psi_{\rm in}}\to \sum_i \hat A_i\proj{\psi_{\rm in}}\hat A_i^\dagger$.
With probability $\bra{\psi_{\rm in}}\hat A^\dagger_i\hat A_i\ket{\psi_{\rm in}}$,
the state is transformed as $\ket{\psi_{\rm in}}\to \hat A_i\ket{\psi_{\rm in}}$.

Let us assume that $N$ parties share a state $\ket{\psi}$ of the form~(\ref{eq:ghz_p}) 
with $q$ such that the state is {\em not} useful, and let us choose 
$\phi=0$ for convenience. Then a single party
could perform the general measurement with the two-outcome measurement
$\hat A_1=\sqrt{1-q}\proj{0}+\sqrt{q}\proj{1}$ and 
$\hat A_2=\sqrt{q}\proj{0}+\sqrt{1-q}\proj{1}$. With probability $P=2q(1-q)$,
the state is transformed into the NOON state, while with probability $1-P$,
the state $\ket{\psi_2}=(q\ket{0}^{\otimes N}+(1-q)\ket{1}^{\otimes N})/\sqrt{1-P}$
is obtained.
Hence in one case, the maximally useful NOON state is obtained, while
in the other case, the state is still as useful as the original state
(namely, shot-noise limited). 
Therefore, the classification of usefulness changes in this situation.
However, from an experimental point of view, 
CLU or LU operations are significantly easier to implement in general.

This result has an implication regarding a possible measure
of entanglement which is useful for sub shot-noise interferometry.
For LU operations $U_{\rm L}$, the quantity 
\be
	e(\rho)=\max\big[0, \max_{U_{\rm L}} F_Q[\rho;\hat J_y] - N\big]
\ee
defined for arbitrary mixed states $\rho$ 
satisfies the following conditions which are typically required of
an entanglement measure \cite{PlenioQIC07,HorodeckiRMP09}: (i) $e(\rho)=0$ for separable states and
(ii) $e(\rho)$ is invariant under LU operations. 
However, the example above shows that it violates the postulate that the
function should not increase on average under LOCC since
\be
	e(\ket{\psi}) < P\cdot e(\ket{{\rm NOON}})+(1-P)\cdot e(\ket{\psi_2})
\ee
holds because we chose the initial state $\ket{\psi}$ such that $e(\ket{\psi})=0$.

\section{Examples}
\label{sec:examples}

\subsection{Symmetric states}
\label{sec:examples_sym}

From Proposition 4 we know that that Observation 1 delivers the optimal 
$F_Q$ for pure symmetric states. Hence the results obtained for the 
three examples of symmetric states in section~\ref{sec:opt_CLU} are already optimal.

\subsection{Singlet states}

Singlet states of $N$ qubits exist if $N$ is even. By definition, these states fulfil (i) $U^{\otimes N}\ket{\psi}=e^{i\phi}\ket{\psi}$ for some phase $\phi$
and (ii) 
$\hat{\vec J}^2\ket{\psi}=0$. It follows that 
$F_Q[\ket{\psi};\hat J_{\vec n}]=0$ holds for any
direction ${\vec n}$. Hence there is no CLU operations which 
makes these states useful for SSN interferometry with a Mach-Zehnder 
interferometer. Therefore it is natural to ask whether 
or not they can be made useful with LU operations.
This situation can only be achieved with bosons or fermions
which can be individually adressed. 
In the case of Fermions occupying just two modes, 
the only entangled state which can occur in the particle
picture is the two-particle
singlet state 
$\frac{1}{\sqrt{2}}(\ket{01}-\ket{10})$,
which is not useful for interferometry under CLU operations.
An obvious example of a $N$ qubit singlet state of individually
adressable particles is the tensor product of $N/2$ two-qubit singlet states. 
For these states we already know that they can be made useful with LU
operations from Theorem 5. 

In the following, we consider the non-trivial family of $N$ qubit singlet states 
defined in Ref.~\cite{CabelloPRA03} as
\be
	\ket{{\cal S}^{(2)}_N}=\frac{1}{\frac{N}{2}!\sqrt{\frac{N}{2}+1}} \sum_{\cal P}z!(\frac{N}{2}-z)!(-1)^{N/2-z}{\cal P}[\ket{01}^{\otimes N/2}],	
	\label{eq:CabelloSinglets}
\ee
where the sum runs over all permutations of the state 
$\ket{01}^{\otimes N/2}$ and $z$ is the number of $0$'s
in the first $\frac{N}{2}$ positions. Apart from (i) and (ii) they are
(iii) multipartite entangled, (iv) invariant 
under the permutations 
${\cal P}_{ij}\ket{{\cal S}^{(2)}_N}=\ket{{\cal S}^{(2)}_N}$
if $i,j\in [1,...,\frac{N}{2}]$ or $i,j\in [\frac{N}{2}+1,...,N]$,
and (v) invariant up to the factor $(-1)^{N/2}$ under exchange of the first
$\frac{N}{2}$ qubits and the second $\frac{N}{2}$ qubits.
Due to the symmetries (iv) and (v), the covariance matrix of the states has the form
\be
	\gamma^{\rm Singlet}_R=\left(
	\begin{array}{cc}
		\tilde A & \tilde C\\
		\tilde C & \tilde A
	\end{array}
	\right),
	\label{eq:gammaSinglet}
\ee
where $\tilde A$ is a block matrix of the form of Eq.~(\ref{eq:gammaS}),
and $\tilde C$ is a block matrix of $3\times 3$ matrices $C$.
Hence we have to compute the matrices $A,B$, and $C$. Due to (i),
the single particle reduced states fulfil $U\rho^{(k)}U^\dagger=\rho^{(k)}$
for any unitary operation $U$. The only state of a single qubit with
that property is $\Eins/2$, from which ${\vec s}={\vec 0}$ and $A=\Eins$ 
follows. Also due to (i), all reduced two-particle states fulfil
$U\otimes U \rho^{(k,j)}U^\dagger\otimes U^\dagger=\rho^{(k,j)}$.
The only states of two qubits with that property are 
the so-called Werner states \cite{WernerPRA89}
\be
	\rho^{(k,l)}=f\proj{\psi^-}+(1-f)\frac{\Eins-\proj{\psi^-}}{3},
\ee
where $f\in[0,1]$. The two-qubit singlet state is the only pure state of 
two qubits fulfilling (i). It follows that 
\be
	\lambda^{(k,l)}=\left(
	\begin{array}{cc}
		1 & {\vec 0}^T\\
		{\vec 0} & \xi\Eins
	\end{array}
	\right),
\ee
where $\xi=\frac{1}{3}-\frac{2}{3}f$ and the matrix is defined as in Eq.~(\ref{eq:lambda}). 
Hence the matrices $A,B$, and $C$ are proportional to the identity.
For reduced two-particle states within 
the sets considered in (iv), $f=0$ has to hold since the reduced state is acting 
on the symmetric subspace only, and we obtain $B=\frac{1}{3}\Eins$.
The missing parameter from $C$ can be calculated by employing (ii) since
it implies $\gamma_C=\hat 0$, where $\gamma_C$ is the collective covariance
matrix introduced in Observation 1. Setting $C=\xi_C\Eins$, 
we find that this condition is fulfilled provided that
\be
	\xi_C=-\frac{2}{N^2}\Big[\frac{2}{3}N+\frac{N^2}{6}\Big].
\ee
We can find the optimal directions $\{{\vec n}^{(k)}\}$ using Observation 2
by diagonalizing $\gamma^{\rm Singlet}_R$ and showing that the maximal eigenvector
has the properties of ${\vec m}$ from Eq.~(\ref{eq:nonsymproblem}). 
As expected, we find $\gamma^{\rm Singlet}_R {\vec m}^{(+)}= 0$ for 
symmetric eigenvectors $({\vec m}^{(+)})^T=({\vec n}^T,{\vec n}^T,...,{\vec n}^T)$,
while the vectors $({\vec m}^{(-)})^T=({\vec n}^T,{\vec n}^T,...,-{\vec n}^T,...,-{\vec n}^T)$
are eigenvectors with eigenvalue $\frac{4}{3}+\frac{1}{3}N$.
In ${\vec m}^{(-)}$, the vectors for 
the particles $\frac{N}{2}+1,...,N$ have the minus sign. 
Finally, the eigenvalue 
$\frac{2}{3}$ is shared by the vectors ${\vec x}_k^{(1)}$, which have vanishing 
elements except for a vector ${\vec n}$ at the positions of particle $1$ and
a vector $-{\vec n}$ at the entries of particle $k\in[2,\frac{N}{2}]$, and the vectors ${\vec x}_k^{(2)}$,
which have vanishing elements except for a vector ${\vec n}$ at the positions of party $\frac{N}{2}+1$ and
a vector $-{\vec n}$ at the entries of party $k\in[\frac{N}{2}+2,N]$.
We did not further specify the vectors ${\vec n}$ because they are eigenvalues of 
the identity matrix in three dimensions since $A,B,$ and $C$ are proportional to $\Eins$.

Hence the vectors ${\vec m}^{(-)}$ are the eigenvectors with the maximal 
eigenvalue,
and we conclude that they lead to the maximal quantum Fisher information
\be
F_Q^{\rm max}({\cal S}_N^{(2)})=\frac{N^2}{3}+\frac{4}{3}N
\ee
surpassing the shot-noise limit
for all $N$. This bound can be reached by keeping $\ket{{\cal S}_N^{(2)}}$
unchanged while choosing the collective operator $\hat J'$ such that
$-\hat\sigma_y^{(k)}$ for the particles $k=1,...,\frac{N}{2}$, 
and $\hat\sigma_y^{(k)}$ for the remaining
ones, for instance. If we consider instead a LU applied to the initial state
and the Mach-Zehnder operator $\hat J_y$, then we can apply $\hat\sigma_z$ to the first $\frac{N}{2}$
parties only. Due to the definition of $z$ in Eq.~(\ref{eq:CabelloSinglets}), 
$\frac{N}{2}-z$ is the number of $1$'s in the first $\frac{N}{2}$ positions.
So the effect of this LU transformation is to remove 
the factor $(-1)^{N/2-z}$.

Any singlet state of $N$ ($N$ even) qubits can be obtained from 
superpositions of permutations of tensor-products of two-qubit singlet states.
It is an interesting question whether or not {\em all} such states
can be made useful for SSN interferometry with LU operations.
The usefulness of singlet states in the mode-picture has been 
considered recently in a different scenario in Ref.~\cite{CablePRL10}.
\\[3mm]

\subsection{Graph states}

Finally, we discuss the usefulness of the so-called 
graph states of $N$ qubits, which recently have received 
large attention because of their importance for 
one-way quantum computation, quantum error correcting codes,
studies of non-locality, and decoherence (see \cite{HeinProc05}
and references within). After discussing general properties 
of graph states in relation to the usefulness for SSN 
interferometry, we consider the so-called cluster states 
and again the NOON state (usually referred to as the GHZ state \cite{GreenbergerAJP90}
in this context).

\begin{figure}[t]
 \includegraphics[width=0.43\textwidth]{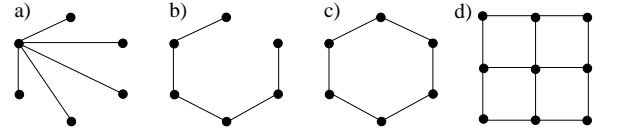}
\caption{Examples for graphs describing important graph states. 
a) describes a GHZ or NOON state (up to LU operations) 
\cite{HeinPRA04,DuerPRL03}, b) shows a linear cluster graph, c) shows a
 ring cluster graph, and d) shows a cluster graph in two dimensions. } \label{fig:graphs}
\end{figure}

Let us first recall the definition of graph states.
A graph $G$ is a collection of $N$ vertices and 
connections between them, which are called edges
\cite{HeinProc05}.
In a physical implementation, the vertices correspond to 
qubits and the edges record interactions (to be specified below) 
that have taken place between the qubits. For each vertex $i$ we define the neighborhood 
$N(i)$, the set of vertices connected by an edge with $i$,
and associate to it a stabilizing operator
\be
	\hat K_i=\hat\sigma_x^{(i)}\bigotimes_{j\in N(i)}\hat\sigma_z^{(j)}.
\ee
It is easy to see that all the stabilizing operators commute.
The graph state $\ket{G}$ associated to the graph $G$ is the 
unique $N$-qubit state fulfilling
\be
	\hat K_i\ket{G}=\ket{G}\quad {\rm for}\ i=1,2,...,N.
\ee
From a physical point of view, one can also define a graph state 
as the state arising from $[(\ket{0}+\ket{1})/\sqrt{2}]^{\otimes N},$ 
if between all connected qubits $i,j$ the Ising-type interaction 
$\hat H_I= (\eins - \sigma_z^{(i)})\otimes (\eins - \sigma_z^{(j)})$
acts for the time $t=\pi/4$. See Fig.~\ref{fig:graphs} for examples 
of prominent graph states.

The group of products of the $\hat K_i$ is called stabilizer ${\cal S}$
\cite{GottesmanPRA96}. 
The state $\ket{G}$ can be expressed with the 
elements of the stabilizer \cite{HeinPRA04,DuerPRL03},
\be
	\proj{G}=\frac{1}{2^N}\sum_{s\in {\cal S}} s.
	\label{eq:Gstabilizer}
\ee
This form is particularly useful for our purpose, because it allows to read off directly 
the  reduced one- and two-qubit density matrices in the form of Eq.~(\ref{eq:rhored}): 
\\[1mm]
\noindent
{\bf Observation 6}. (i) For the reduced state of $p$ qubits, products 
of at most $p$ stabilizers $\hat K_i$ contribute. (ii) For $p=1$, the reduced state
is $\rho^{(r)}_i=\frac{1}{2}\Eins$ unless qubit $i$ is not connected to any other 
qubit, in which case $\rho^{(r)}_i=\frac{1}{2}(\Eins+\hat\sigma_x^{(i)})$. 
(iii) If $\rho_{ij}$ is a reduced state of $p=2$ qubits, then the stabilizers $\hat K_i$ (or $\hat K_j$) contribute if $i$ is the
only neighbor of $j$ (or vice versa). Also, the products
$\hat K_i\hat K_j$ contribute if the qubits $i$ and $j$ have the
same neighbors, where
it is irrelevant whether or not $i$ and $j$ are neighbors themselves.
\\[1mm]
\noindent
{\em Proof.} All elements $s$ of the stabilizer have the form 
$\pm\otimes_{i=1}^N\hat\sigma^{(i)}_{j^{(i)}}$.
Hence $\tr_i[s]=0$ unless $j^{(i)}=0$ (since then $\hat\sigma^{(i)}_{j^{(i)}}=\Eins$) 
and in analogy if more 
than one qubit are traced out \cite{HeinProc05}. 
If we compute the reduced one- and two-qubit density matrices from 
Eq.~(\ref{eq:Gstabilizer}), then only those $s$ will contribute
which act as the identity on the traced out particles.
More specifically, products of $\tilde p$ stabilizers are of the form 
\be
	\prod_{k=1}^{\tilde p} \hat K_{i_k} \propto \Big(\bigotimes_{k=1}^{\tilde p} \hat\sigma_x^{(i_k)}\Big)
	\Big(\bigotimes_{k=1}^{\tilde p} \big(\otimes_{j_k\in N(i_k)}\hat\sigma_z^{(j_k)}\big)\Big).
\ee
Since $[\hat K_i,\hat K_j]=0$ and $\hat K_i^2=\Eins$ for all $i$ and $j$,
we only have to consider products of different stabilizers for a given $\tilde p$.
Then (i) follows because if ${\tilde p}$ is larger than the number of qubits $p$
in the reduced state, one or more $\hat\sigma_x$ operators remain
acting on the rest, which remain traceless even when multiplied by the $\hat\sigma_z$ 
operators acting on the neighborhoods. (ii) and the first part of (iii) follow directly,
while the second part of (iii) follows because otherwise $\hat\sigma_z$ 
operators would be left acting on qubits which are traced out. \proofend

Since we are not interested in the situation where a qubit is fully separable 
from the rest, we can assume that ${\vec s}^{(k)}=\vec 0$ for any $k$
in the following, since the reduced states are equal to $\frac{1}{2}\Eins$ 
in this case. From Observation 6 it follows directly that cluster states 
of all kinds are practically of no more use for SSN interferometry than product states:
\\[1mm]
\noindent
{\bf Proposition 7}. 
The maximal quantum Fisher information $F_Q^{\rm max}$
of linear cluster states with $N \geq 4$
particles is $N+4$. For $N \geq 5$ qubits, ring cluster states as well
cluster states in more than one dimension have $F_Q^{\rm max}= N.$
\\[1mm]
\noindent
{\em Proof.} For  $N=3$ the linear cluster states is LU equivalent to
a GHZ state  and for $N=4$ the ring cluster state is equivalent to a 
linear cluster state \cite{HeinProc05}. The claim for the ring cluster 
state and the cluster states in more than one dimension follows 
directly from Observation 6, as all reduced two-qubit density matrices are 
of the form $\rho^{(r)}=\frac{1}{4}\Eins$, and hence 
$\gamma_R=\Eins$. Then Observation 2 yields $F_Q^{\rm max}\le N$.
For linear cluster states, there are four off-diagonal elements
of $\gamma_R$ coming from the stabilizers $\hat K_1=\hat\sigma_x^{(1)}\otimes\hat\sigma_z^{(2)}$
and $\hat K_N=\hat\sigma_z^{(N-1)}\otimes\hat\sigma_x^{(N)}$
at the ends of the cluster.
Writing down $\gamma_R$ in the block-order $1,2,N-1,N,3,4,...,N-2$ yields
\be
\gamma_R=\left(
		\begin{array}{cc}
			\Eins & \hat x\hat z^T \\
			\hat z\hat x^T & \Eins
		\end{array}
	\right)
	\oplus
	\left(
		\begin{array}{cc}
			\Eins & \hat z\hat x^T \\
			\hat x\hat z^T & \Eins
		\end{array}
	\right)
	\oplus\Eins;
\ee
where $\hat x^T=(1,0,0)$ and $\hat z^T=(0,0,1).$
This matrix can be reordered as $\gamma_R=(2\ketbra{x+})\oplus(2\ketbra{x+})\oplus\Eins_{N-4}$,
where $\ket{x+}=\frac{1}{\sqrt{2}}\binom{1}{1}$. Hence from Observation 2 it follows 
that $F_Q^{\rm max}\le 2N$. However, this limit cannot be reached due to the 
restriction on ${\vec m}$. Due to the block-diagonal structure of $\gamma_R$,
the largest expectation value is obtained by choosing 
${\vec m}^T=(\hat x^T,\hat z^T,{\vec n},...,{\vec n},\hat z^T,\hat x^T)$, 
where ${\vec n}$ may point in any direction, which leads to $F_Q=N+4$.
\proofend

A phase estimation scheme for one-dimensional cluster states
enabling SSN sensitivity was suggested in Ref.~\cite{RosenkranzPRA09}. 
In contrast to the situation we considered, the authors suggested to use
superpositions of cluster states and a non-collective generator 
of the phase shift, showing
that in this case SSN sensitivity is possible even in the presence of noise.

Let us finally illustrate why the GHZ states have the largest Fisher information
possible from the point of view of graph states. 
From Fig.~\ref{fig:graphs} a) we see that if the qubit 1 is
connected to all the others then all $\hat K_i$, $i\neq 1$ 
contribute to the reduced two-qubit states since all those qubits
have only one neighbor. Further,
all the products $\hat K_i\hat K_j$ for $i,j\neq 1$ contribute since
these qubits have all the same neighborhood.
The covariance matrix $\gamma_R$ then takes the form
\be
	\gamma_R=\left(
		\begin{array}{ccccc}
			\Eins & \hat z\hat x^T & \hat z\hat x^T & \cdots & \hat z\hat x^T\\
			\hat x\hat z^T & \Eins & \hat x\hat x^T & \cdots &  \hat x\hat x^T\\
			\vdots & \vdots & \vdots &\vdots & \vdots\\
			\hat x\hat z^T & \hat x\hat x^T &\hat x\hat x^T & \cdots &\Eins
		\end{array}
	\right).
\ee
It can be directly checked that 
${\vec m}^T=\frac{1}{\sqrt{N}}(\hat z^T,\hat x^T,\hat x^T,...,\hat x^T,\hat z^T)$
is the eigenvector of $\gamma_R$ corresponding to the maximal eigenvalue $N$,
and hence $F_Q^{\rm max}=N^2$.

\section{Conclusion}
\label{sec:conclook}

We have studied linear two-mode interferometers from a quantum information
theory perspective. 
In particular, we have adressed the question of whether or not all pure entangled 
states of $N$ particles can achieve sub shot-noise sensitivity 
in such interferometers if they can be optimized by 
operations which are local in the particles. 
We used the Cram{\'e}r-Rao
theorem, which gives a lower bound on the optimal sensitivity 
{\em via} the quantum Fisher information $F_Q$. For $F_Q>N$,
sub shot-noise sensitivity can be achieved in the central limit.

We have studied the maximal quantum Fisher information $F_Q$ 
that can be achieved for a general two-state linear 
interferometer such as the Mach-Zehnder interferometer.
We have found a simple way to determine the optimal CLU operation, 
and an upper bound for the optimal $F_Q$ for LU operations,
which is tight in many cases.
The optimizations carry over directly to the mixed state case
and are useful for the experimental optimization of the source
if tomographic data of the state is available.

Using these results, we have fully characterized the pure symmetric entangled states
which are of no more use than non-correlated states under CLU operations.
These states and operations are available in bosonic two-mode interferometers.
Further, we have obtained that for symmetric states of particles which
can be individually adressed, a CLU operations achieves the maximal
$F_Q$ even if arbitrary LU can be applied. From these
results it follows that while for $N=2$ any entangled state can 
be made useful with LU operations, there are pure entangled states, 
and even fully $N$-partite entangled states, which are not useful
for sub shot-noise interferometry. We briefly commented that this
picture changes when more general local operations are available.

Finally, we discussed several interesting states from the literature, finding
the optimal sensitivity that they can deliver. Among them, we find that the 
highly entangled cluster states, which comprise a resource for one-way
quantum computation \cite{HeinProc05}, are practically not more useful than separable states.

We thank J.I. Cirac for fruitful discussions and G. T{\'o}th for pointing 
out reference \cite{ChenPRA05} to us. OG acknowledges support by the FWF 
(START prize and SFB FoQuS) and the EU (NAMEQUAM, QICS, SCALA).


\begin{thebibliography}{51}
\bibitem{Note_indist}
It has been argued that for indistinguishable particles, the
entanglement defined in this way is not physical since the individual particles cannot
be adressed \cite{ZanardiPRA02,BenattiAoP10}. Nontheless, in the context of interferometry, 
it is a useful resource for sub shot-noise sensitivity.
\bibitem{Note_Obs1}
The conversion from ${\vec n}_{\rm max}$ to $U_{\rm CL}^{\rm opt}=\otimes_{k=1}^N U^{\rm opt}$ 
can be done as follows.
If ${\vec n}_{\rm max}=O^T\hat y$ has been found, then from ${\hat y}\cdot{\vec n}_{\rm max}=\cos(\phi)$
and ${\hat y}\times {\vec n}_{\rm max}=\sin(\phi)\ \hat n$ the angle $\phi$ and the axis of 
rotation $\hat n$ can be extracted, and 
$U^{\rm opt}=\exp(i\phi \hat{\vec\sigma}\cdot{\hat n}/2)$
is (up to a phase) the unitary matrix such that 
$(U^{\rm opt})^\dagger \hat\sigma_{\hat y} U^{\rm opt}=\hat\sigma_{{\vec n}_{\rm max}}$ \cite{SakuraiBook}. 
The positive sign in the exponent is due to the fact that $O^T$ 
appears in the definition of ${\vec n}_{\rm max}$.
\bibitem{Note_Obs1b}
To see this, note that Eq.~(30d) in Ref.~\cite{TKGB09}
implies that for separable states 
$
\lambda_{\rm max}(N \gamma) \leq (N-1) Tr(\gamma) - N (N-2)/4
$
since 
$\mathfrak{X} = N \gamma + \mean{\hat {\vec J}} \mean{\hat{\vec J}}^T \geq N \gamma.$
Using Eq.~(30b) implies then that
$
N \lambda_{\rm max}(\gamma) \leq N (N-1)/2  - N (N-2)/4 
$
hence
$
\lambda_{\rm max}(\gamma)\leq N/4.
$
We used the notations of Ref.~\cite{TKGB09} here.
\end{thebibliography}

\end{document}